 \documentclass[12pt,preprint]{aastex}

\begin{document}

\title{Spectral Lags Explained as Scattering from Accelerated Scatterers}
\author{ David Eichler and Hadar Manis}

\affil{Department of Physics, Ben-Gurion University, Beer-Sheva,
Israel, 84105;
  eichler@bgu.ac.il}
\begin{abstract}
A quantitative theory of spectral lags for  $\gamma$-ray bursts
(GRBs) is given. The underlying hypothesis is that GRB subpulses are
photons that are scattered into our line of sight by local
concentrations of baryons that are accelerated by radiation
pressure. For primary spectra that are power laws with exponential
cutoffs, the width of the pulse and its fast rise, slow decay
asymmetry is found to increase with decreasing photon energy, and
the width near the exponential cutoff  scales approximately as
$E_{ph}^{-\eta}$, with $\eta\sim 0.4$, as observed. The spectral lag
time is naturally inversely proportional to luminosity, all else
being equal, also as observed.
\end{abstract}
\keywords{ gamma-rays: bursts}


\section{Introduction} The fast rise, slow decay of subpulses in GRB is a common
feature. There could be many ways to explain it (e.g. impulsive
energy infusion followed by slower cooling or light echoing).  It is
therefore desirable to discriminate among the different models with
quantitative tests and predictions whenever possible.

 In a previous
paper (Eichler and Manis 2007, hereafter EM07), it was
 suggested that fast
 rise,
 slow decay subpulses constitute a qualitative manifestation of
 baryons being accelerated by radiation pressure. More generally, the
 basic idea can apply to any flow in which a light, fast fluid imparts
 energy to a clumpy, denser  component of the flow by overtaking
 the clumps from the rear, but for convenience in this discussion  we refer to the
 fast light component as photons that scatter off the clumps.
  It was proposed
 that the fast rise of a subpulse is the stage where a cloud of
 baryons scatters photons into a progressively  narrowing beaming cone of
 width $1/\Gamma$, where $\Gamma$ is the bulk Lorentz factor of the
 accelerating cloud. This narrowing of the $1/\Gamma$ cone causes brightening as
 long as $\Gamma$ remains below $1/\theta$, where $\theta$ is the
 viewing angle offset between the observer's line of sight and the
 velocity vector of the scattering cloud.
Once  the scattering cloud accelerates to a Lorentz factor exceeding
$1/\theta$, the viewer is no longer inside the beaming cone and
apparent luminosity begins to decline. If the cloud accelerates with
roughly constant radiative force, as is reasonable to suppose over
timescales that are short compared to the hydrodynamic expansion
time, then the decline in luminosity is considerably slower than the
rise time, because the acceleration time increases so dramatically
as the velocity approaches c. It was shown in EM07 that the spectral
peak frequency as seen by the observer remains roughly constant
during the rising phase and, well into the declining phase, softens
as $t^{-2/3}$, as reported by Ryde (2004).

The spectral softening of the pulse has been carefully studied by
Norris and coworkers, who have noted that the asymmetry of the
subpulse increases with decreasing frequency and that the width of
the subpulse scales roughly as the frequency to the power -0.4
(Fenimore et al 1995) in the BATSE energy range. This represents
additional information, as the result of Ryde is in principle
consistent with symmetric pulses.

 In this letter, we derive the light curves as a function of both
 time and frequency. We
  show that the asymmetry of the subpulse decreases with frequency and
  that the hypothesis of
EM07 is  quantitatively consistent with the formulation of Fenimore
et al (1995).

The basic assumption in our hypothesis - that a scattering screen
can {\it enhance} the detected signal - presupposes that the
unscattered radiation is beamed and directed slightly away from the
observer's line of sight, so that the scattering of photons into the
line of sight creates a "flash-in-the-pan" type brightening. This
assumption is non-trivial, but has been suggested as being an
explanation for the Amati relation (2002) in earlier papers (Eichler
and Levinson 2004, 2006; Levinson and Eichler 2005). In this series
of papers, it was suggested that a significant fraction of all GRB
are actually brighter and harder in spectrum than they appear to be,
and that they appear dimmer and softer because we, the observers,
are viewing the burst from a slightly offset angle relative to the
direction of the fireball. The interpretation of the subpulses given
here and in EM07 is thus in accord with this picture.

\section{Pulse Profiles at Different Photon Energies}

The equations describing matter that is being accelerated by highly
collimated radiation pressure were presented in EM07. Here we  apply
the solutions described in EM07 to calculate the profile of a
subpulse as a function of photon energy.  We assume that the
differential  primary photon spectrum  $N_p(E)$ has the form
$E_o^{-\alpha}$exp(${-\zeta E_o}$), where $E_o$ is the photon energy
in the frame of the central engine. This form is consistent with a
Comptonized thermal component. It does not, however, exclude a power
law photon spectrum produced further downstream by internal shocks.
After scattering off a baryon clump that moves with velocity
$\bf{\beta} c$, the photon energy as seen by an observer at angle
$\theta$ is

\begin{equation}
E=E_o/[\Gamma^2(1+\beta)(1-\beta cos\theta)]=E_o(1-\beta)/(1-\beta
cos\theta).
\end{equation}
Together with the solution for the accelerating trajectory
$\beta(t)$ given in EM07, the source/observer frame invariance of
the number of photons $N(E)dEdtd\Omega$  scattered  within energy
interval dE and time interval dt, and solid angle $d\Omega$,
equation (1) determines the light curve  N(E,t)
 as a function of observed photon energy E and  observer time t.

In figure 1 the subpulse light curves are plotted for three
different frequencies.  It is clear that the pulse width is larger
and the rise-fall asymmetry is more pronounced at lower frequencies,
as reported by Fenimore et al. (1995) and references therein. In
figure 2 the width is plotted as a function of photon energy. At
high energies, which correspond to the BATSE measurements used by
these authors, the width is seen to scale approximately as the
photon energy to the power $-0.4$, as reported by Fenimore et al.,
above $10^2$ KeV. Similar calculations with varying values for the
low energy power law index, $\alpha$,  of the primary spectrum show
that this dependence  is  weakly dependent on $\alpha$ and on
viewing angle. For a viewing offset angle of 10  degrees, the width
depends on $E^{-\eta}$, with $0.4 \le \eta \le 0.5$ when $-0.75 \le
\alpha \le 0$ with the sensitivity $d\eta/d\alpha \sim 0.08$ at
$\alpha=-0.7$. For viewing offset of 15 degrees, the value of $\eta$
is increased by about 0.033 so that a given range of $\eta$ is
occupied by a somewhat lower (i.e. more negative) range of $\alpha$
than for smaller viewing offsets. For an extended beam, some
contribution from larger offsets is inevitable, but a synthesis of
light curves from extended beams is deferred for future work.  It
can be seen from figure 2 that the value of $\eta$ increases with
$\zeta E$, and the range of $\zeta E$ that corresponds to BATSE
sensitivity depends on cosmological redshift, larger z implies
larger intrinsic values of $\zeta E$, hence steeper E dependence of
the pulse width, over a given range of observed photon energies.
Finally, the primary source spectrum, which we argue is not a direct
observable, is somewhat uncertain. Altogether, the range of $0.4 \le
\eta \le 0.5$ is consistent with the ranges $-1 \le \alpha \le 0$,
$1\le z \le 3$, $0.5  \le \zeta E \le 2$, and $0.10 \le \theta \le
0.25$. It is predicted that the dependence  of width on E weakens
(i.e. $\eta$ decreases) at lower photon energies, and this should be
testable with detectors that are more sensitive at lower energies,
such as the Gamma Ray Burst Monitor.

 As the acceleration time is inversely proportional to the
 radiation flux on the scatterer, it is clear, all other things
 being equal, that the rise time of the pulse and spectral lag are
 inversely proportional to source luminosity, as observed (e.g.
 Gehrels et al., 2006). Of course, scatter in other variables,
 such as the  distance from the
 source of illumination, optical depth of the scatterer etc.,
   creates scatter in the constant of
 proportionality.

If the scattering is isotropic (or backwardly biased due to high
optical depth)  in the scattering frame, it follows from equation
(1) that the scattered radiation, averaged over angle, is a factor
of 2 (or more) softer than the primary emission. As explained in
EM07, the other half of the energy goes into the acceleration of the
scatterer. On the other hand, at most viewing angles $\theta$, the
scattered radiation is harder than scattered radiation after the
scatterer has reached terminal Lorentz factor $\Gamma_f$ if
$\Gamma_f \ge 1/\theta$. As the scattered radiation during the
acceleration phase of the scatterer is likely to be the most time
dependent, it may be possible to separate out this component from
the other two. The extent to which the scattering affects the
spectrum depends, of course, on the fraction of primary radiation
that is scattered. Equating the scattered photon energy with the
baryon afterglow energy, and applying the results of Eichler and
Jontof-Hutter (2005), which estimated the afterglow efficiency, we
may tentatively estimate that about 30 percent of the primary
emission is scattered, about half of that 30 percent going into
baryons and the other half
 ending up in a scattered subpulse component. Clearly
there is variation in the scattered fraction as well as uncertainty
in theoretical inferences of the baryon energy from afterglow
calorimetry, so this estimate should be considered rough and
preliminary.

 The time-integrated spectrum at a {\it given} viewing angle can
 be different from the
  average, because the scattered radiation is not isotropic but,
 rather, beamed in an ever narrowing cone as the scatterer
 accelerates. Consider a primary emission spectrum that is a delta
 function $\delta(E - E_o)$.
At a given  $\mu\equiv cos\theta$ and a given observed photon energy
$E$, a monochromatic primary spectrum $\delta( E-E_o)$ is, after
scattering,  monochromatic at photon energy $E[\beta(t)]$  given by
equation (1), so the contribution to the emitted power at energy E
comes only at
\begin{equation}
\beta(t)=\frac{1-E/E_o}{1-\mu E/E_o}
\end{equation}
 The time integrated energy  $d^2F(E,\theta)/dEd\Omega$ of the scattered
 radiation at  observed
 photon energy E and viewing angle $\theta$ is $d^2F/d\Omega dE = d(\int
 (dP/d\Omega)dt)/dE$ = $(dP/d\Omega)
 (d\beta/dt)^{-1}
 (-dE/d\beta)^{-1}$ where $P(E, \theta)$ is
 the  power of the  scattered radiation as observed at
  photon energy E in the
 frame of the primary source,\footnote{P has
 units of power rather than power per unit energy as
  the primary spectrum is taken here to be monochromatic. The minus sign
  in front of $dE/d\beta$ is to make it a positive quantity.} t is
 the elapsed time in the frame of
 the source (and in any case the  variable of integration),
 $d\beta/dt$ is\footnote{This expression is for an optically thin source, in the approximation
 that the primary radiation is radially combed, and under the
  assumption that the scattered radiation
 $dP^{\prime}/d\Omega^{\prime}$  has front-back
 symmetry in the frame of the scatterer. A high,
 time varying optical depth would be more complicated.}
 $\frac{\sigma_T}{m_p}\Gamma^{-3}P^{\prime}/c$ (EM07). Making the
 simplifying assumption that the scattered radiation is isotropic in
 the frame of the scatterer, i.e. that
 $dP^{\prime}/d\Omega^{\prime}=P^{\prime}/4\pi$, using the
 transformation for emitted power
 $dP/d\Omega = \Gamma^{-4}(1-\beta \mu)^{-3}dP^{\prime}/d\Omega^{\prime}$ (equation  4.97a in Rybicki and Lightman,
 1979)
  and evaluating $dE/d\beta$
    in units of $E_o$
    from equation (1), it follows that
 \begin{equation}
 d^2F/d\Omega dE
=[(E/E_o)(2-E/E_o)(1+\mu)]^{1/2}/E_o[1-\mu]^{3/2}.
\end{equation}
In the limit that $\mu$ is sufficiently below
 unity that $(1-\mu E)$ does not depend significantly on E,
$ d^2F/d\Omega dE$ is proportional to $E^{1/2}$, i.e. $\alpha =
-0.5$.  That this is softer than the average over all viewing
angles - for which $\alpha=0$ -
can be understood as the result of most of the emission at large t
 going into a narrower cone than the one the observer is on,
so that the observer sees only the soft fringes of this dominant
component. Also note that this result assumes that the scatterer's
acceleration proceeds indefinitely. If the scatterer reaches a
terminal velocity $\beta_a c$, then the observer would not see any
of the primary radiation originally at $E_o$ scattered to an energy
below $E=E_o(1-\beta_a)/(1-\beta_a cos\theta)$.

 While the scattered radiation is not the only
observable component, the hypothesis that it comprises a
 significant
fraction of the total fluence of many GRB is broadly consistent with
the tendency of the low energy spectral index to not exceed 0. The
question of how much radiation is scattered, before and after the
scatterers have reached terminal velocity, remains somewhat open at
the quantitative level, but the  GBM data may provide the
opportunity to  address these questions
 quantitatively as well as qualitatively.

\section{Discussion and Conclusions}

We have presented evidence that subpulses in $\gamma$-ray bursts are
photons that are scattered into our line of sight by scatterers with
lower Lorentz factors than the frame in which the prescattered
photons had zero net momentum.  Because scattering  can never
increase the intensity of a beam of photons, the hypothesis
presupposes that the observed intensity is lower in the observer's
direction than in the direction of the beam, and that the scattering
by the slower baryons broadens the beam enough that it engulfs the
observer's line of sight. It is this widening of the beam that
allows the observer to see enhanced flux. This fits the picture
already put forth to explain the Amati relation.

As most GRB with known redshifts have spectral peaks and energies
that are both below those of the hardest, brightest GRB, it would
follow, according to our interpretation, that a large fraction of,
perhaps most, GRB are observed from an offset viewing angle. The
question then arises as to why there are so many offset observers
relative to those that are within the $1/\Gamma_{ph}$ beaming cone
of the primary radiation. In an earlier paper (Eichler and Levinson
2004), it was proposed that the complex shape of the primary beam -
e.g. an annular shape - would allow a comparable number of  viewers
just off (i.e. within several times $1/\Gamma_{ph}$)  of  the
periphery of the primary beam to the number of viewers within this
periphery. It was shown that a thick annulus, in which the
separation between the inner and outer radii is comparable to half
the outer radius, gives a distribution of offsets that is consistent
with the observed distribution of spectral peaks. There may be
several reasons for GRB jets to have an annular or otherwise complex
morphology: It could be that baryons are entrained in the flow from
( or are fed neutrons by) the confining walls  (Levinson and Eichler
2003) and that much of the liberated energy is due to dissipation
associated with this entrainment. Or, it could be that dissipation
from shocks associated with wall impact could preferentially
liberate energy from near the confining walls (Begelman, private
communication).

Here we suggest another simple mechanism that would give the
scattered $\gamma$-radiation some measure of  annular bias,
depending on the fraction scattered: Consider the region of flow
where the $\gamma$-rays make their last scattering off baryons
within the flow. If, as seems more likely than not, the baryons are
clumpy, it is likely that the clumps are optically thick, while the
interclump medium is optically thin. In this case the photons are
most likely to make their last scattering off the surface of a
clump. If the clumps are moving more slowly than the primary
fireball, then the photons are most likely to overtake them from the
rear, and, because the clumps are optically thick, the photon is
likely to emerge from its last scattering from the rear end of the
clump. It is then obscured from a viewer who lies along the velocity
vector of the clump, just as sunlight scattering off the moon is
obscured to a viewer on the dark side of the moon. The viewers best
positioned to see the back side illumination are those who see
photons that are emitted backward in the frame of the cloud - i.e.
those that are offset by more than $1/\Gamma$ from the velocity
vector of the clump. This, we suggest, could be a reason so many GRB
are observed from an offset angle of more than $1/\Gamma$. Some
subpulses have such fast rises that they can be interpreted as the
$1/\Gamma$ shadow of an optically thick cloud narrowing  from above
to below the offset angle of the observer, $\theta$, as the cloud
accelerates (EM07). The very sudden rise is then attributed to the
observer emerging from the shadow of the clump.

To summarize, we suggest that spectral lags from long bursts are the
result of high $\Gamma$ radiation (where $\Gamma$ is the Lorentz
factor of the frame in which the radiation is isotropic) impacting
slower baryons from behind, and scattering off them while
accelerating them. As in (EM07), the quantitative agreement with the
data is excellent. The inverse correlation between lag and
luminosity (e.g. Gehrels et al 2006) follows from that between
acceleration time and luminosity.  The assumptions needed to make
the general scheme work are minimal.\footnote{ The result could even
be obtained from  the internal shock model of GRB if the photon
energy of radiation from  the rear end of the accelerating clump
were to scale linearly with the Lorentz factor of the high $\Gamma$
fluid in the frame of the clump. However, in the simplest internal
shock model, where the average electron energy, magnetic field, and
blue shift all scale with $\Gamma$, the spectral peak varies as a
high power of $\Gamma$.} In any case, the minimal conclusion to be
drawn from the quantitative success of this model in explaining
spectral lags of long GRB is that baryons are still undergoing
considerable acceleration by the time the fireball as a whole is
becoming optically thin. Were energy systematically removed from
baryons beyond the photosphere (e.g. because they collided with
other scatterers in their shadow), one would expect negative
spectral lags.

It is predicted that the time integrated spectra of the subpulses
should be slightly softer than the primary emission, and harder than
the emission that is scattered  at terminal Lorentz factor. The
GLAST/Fermi GBM monitor offers the potential opportunity for
unraveling these three components.

For short hard bursts (SHB),  the subpulses are about a factor of 10
to 30 shorter than for long ones and the spectral lags are much
smaller.  This difference  can perhaps  be attributed to the
difference in timescale over which
 the baryons are undergoing
acceleration. For example SHB  are likely to be  observable somewhat
closer to the central engine, being unobscured by the  envelope of a
massive host star, and at a wider angle (Eichler, Guetta, and Manis
2008). If this is indeed the case, then we may be able to see
baryonic clumps at an earlier stage of their acceleration, when the
acceleration time is considerably shorter. There are many unknowns
in this hypothesis - e.g. the optical depth of the baryons, their
point of injection and their covering  factor - on which the
qualitative nature of the subpulses may depend, and future work will
focus on the question of whether reasonable ranges for these
parameters can explain the wide diversity of GRB light curves and
spectra.

\section*{Acknowledgments}
This work was supported by the Joan and Robert Arnow Chair of
Theoretical Astrophysics, the US-Israel Binational Science
Foundation and the Israeli Science Foundation's Center of Excellence
Program.

\clearpage

\begin{figure}
\epsscale{.90} \plotone{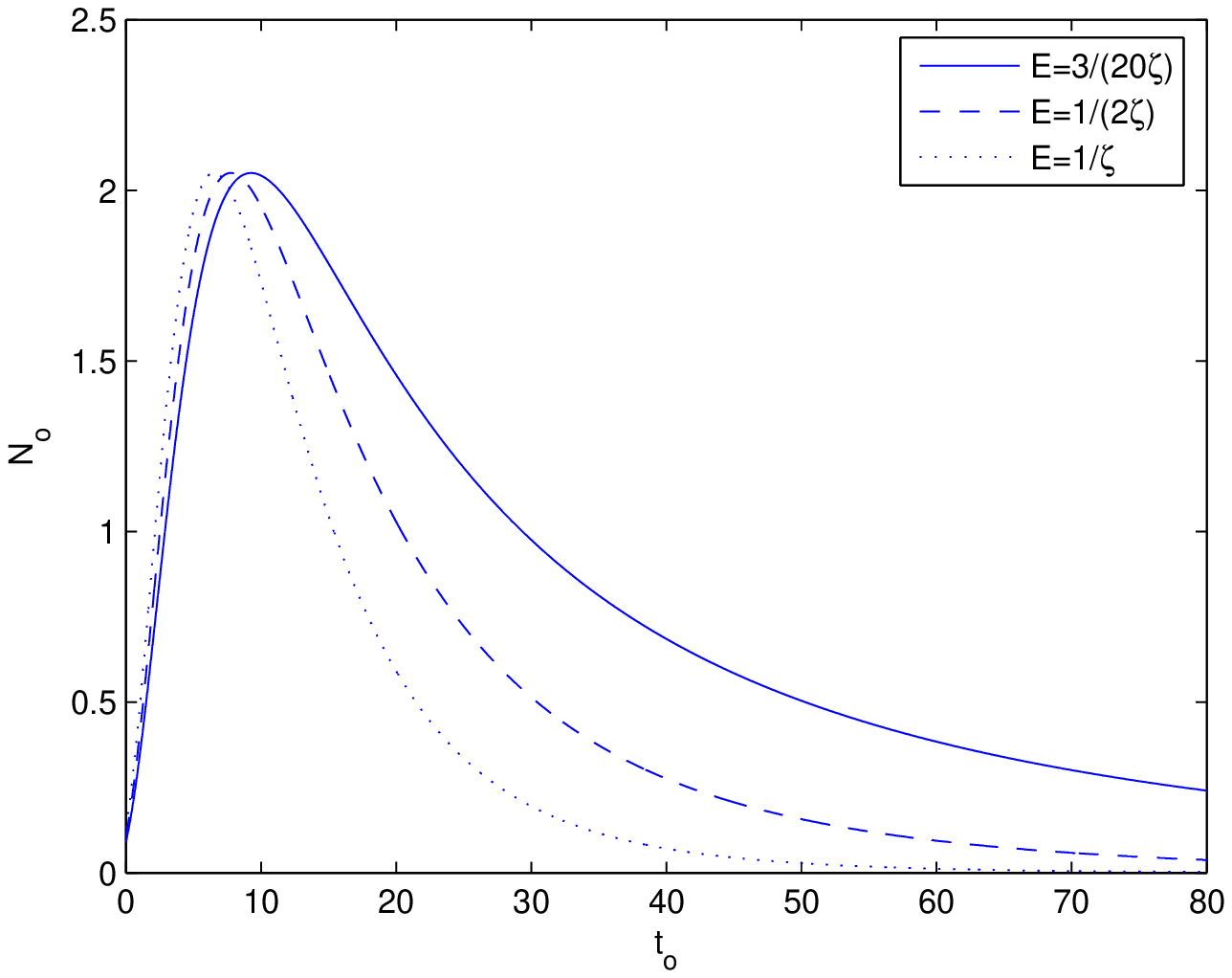} \caption{The predicted profile
of a fast rise, slow decay subpulse for three different frequencies.
The primary spectrum is assumed to have the form
$E_o^{\alpha}e^{-\zeta E_o}$, where $\zeta$ is of order $(600
KeV)^{-1}$.} \label{fig_1}
\end{figure}

\begin{figure}
\epsscale{.90} \plotone{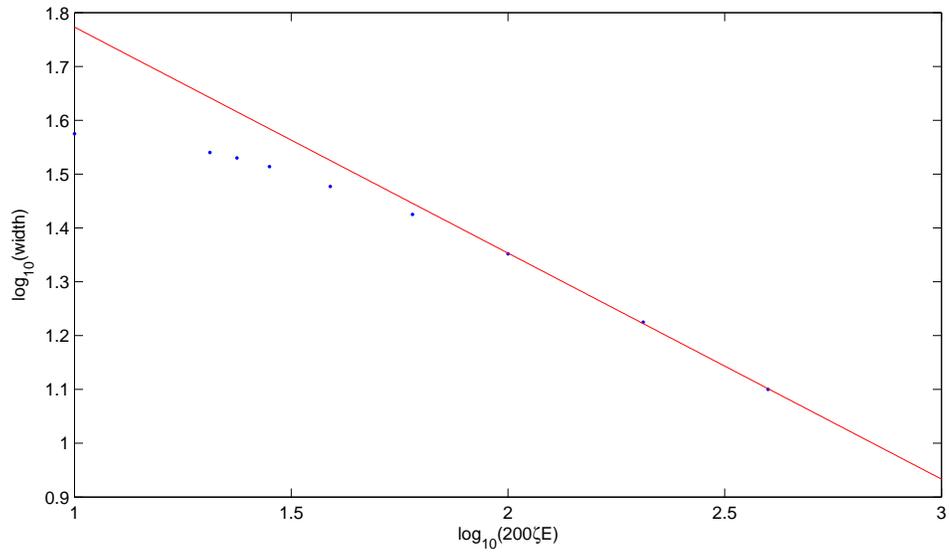} \caption{The pulse width $w$ is
plotted as a function of photon energy E for a primary spectrum
proportional to $E^{-1/2}e^{-\zeta E}$. Here the viewing angle is
taken to be 10 degrees. The red line shows the relation $w \propto
E^{-0.43}$.} \label{fig_2}
\end{figure}

\end{document}